\documentclass[nofootinbib,superscriptaddress]{revtex4-1}

\usepackage{amsmath}
\usepackage{amsfonts}
\usepackage[english]{babel}
\usepackage[latin1]{inputenc}
\usepackage{graphicx}
\usepackage{hyperref}

\begin{document}

\title{Exploring Special Relative Locality with deSitter momentum-space}
\author{Niccol\'{o} Loret}
\email{niccolo@accatagliato.org}
\affiliation{Dipartimento di Matematica, Universit\`a di Roma ``La Sapienza", P.le A. Moro 2, 00185 Roma, Italy\\
and Perimeter Institute for Theoretical Physics, 31 Caroline Street North,
Waterloo, ON, N2L 2Y5 Canada.}

\begin{abstract}
Relative Locality is a recent approach to the quantum-gravity problem which allows to tame nonlocality effects which may rise in some models which try to describe Planck-scale physics. I here explore the effect of Relative Locality on basic special-relativistic phenomena. In particular I study the deformations due to Relative Locality of special-relativistic transformation laws for momenta at all orders in the rapidity parameter $\xi$. I underline how those transformations also define the RL characteristic (momentum-dependent) invariant metric.\\ 
I focus my analysis on the well studied deSitter momentum-space framework and I investigate the differences and similarities between this model and Special Relativity, from the definition of the boost parameter $\gamma$ to a first discussion of  transverse-effects characteristic of Relative Locality on clocks observables.
\tableofcontents
\end{abstract}

\maketitle

\section{Introduction}

Relative Locality (RL) is a quite young approach to the quantum-gravity problem which formalizes 
nonlocalities and other characteristic features of deformed symmetries models, introducing some sort of momentum space curvature \cite{principle,grf2nd,FlaGiu,lateshift} that influences the localization process, at a characteristic scale that we assume to be of the order of the Planck-scale $\ell\sim 1/M_P$ \footnote{In this paper lengths will have the dimensions of an inverse mass, since from now on we will adopt the natural units system $c=\hbar=1$.}.  
A strong motivation to explore this feature has emerged both from the theoretical and the phenomenological sides, since the Planck-scale curvature of momentum space introduces corrections to the travel times of particles, opening also an opportunity for experimental tests \cite{phenomenology}.\\
So far many aspects of this theory have been examined: from the implications for interaction vertices conservation laws \cite{trevisan,grf2nd,FlaZara}, to some attempts to generalise Relative Locality to curve space-time scenarios \cite{KowaJack1,KowaJack2}. However in literature we are still lacking a clear explanation of the properties of the theory transformation laws as deformation of Lorentz ones. Though the argument has been analysed so from the algebraical point of view \cite{majidCURVATURE,jurekDSMOMENTUM,gacmaj}, as from the phenomenological one \cite{trasfallord2,synchrotron}. In this paper we will discuss analogies and divergences between Special Relativity and its Relative Locality version, using the well studied deSitter momentum-space formalism \cite{FlaGiu,lateshift} at first order in the deformation parameter $\ell$. In order to give a satisfying characterization to the Relative Locality features that we will encounter, we will need to work with boost transformations at all orders in the rapidity parameter $\xi$, in 2+1 dimensions (the 3+1D generalisation is straightforward). We will then give a brief description of the RL-boost transverse effects \cite{transverse,transverseproc} on momenta, showing also how those $\ell$-deformed transformations naturally implement a Rainbow metric formalism \cite{SmolinRainbow} for the invariant line-element.\\
A key element of our analysis is the non trivial coordinate system, defined in analogy with deSitter space-time conserved charges $\Pi_0=p_0-H x^k p_k$, $\Pi_i=p_i$ \cite{AntoninoGiuliaDeSitter,lateshift}. These coordinates satisfy the following non trivial Poisson brackets:  
\begin{equation}
\{\chi^i,\chi^0\}=\ell\chi^i\label{noncomm}\,,
\end{equation}
where in our case the index $i$ can assume the values $i=L,T$ (longitudinal and transverse direction). The reason why this coordinatization is more suitable for this kind of discussion, as we will explain later in detail, is that the wordline expression in $\chi^\alpha$ coordinates is momentum-independent \cite{kbob,lateshift}, and therefore we do not encounter any theoretical problem in fixing a reflexive, symmetric and transitive definition for a time-interval. Relative Locality effects on clocks observables in 2+1D will then be discussed  at the end of the paper. 

\subsection{About deSitter momentum-space}

Our mathematical formalism is based on deSitter momentum-space in 2+1 dimensions whose metric is
\begin{equation}
\tilde{\eta}^{\alpha\beta}(p)=\left(
\begin{array}{ccc}
1 & 0 & 0\\
0 & -(1+2\ell p_0) & 0\\
0 & 0 & -(1+2\ell p_0) \\
\end{array}
\right)\,.\label{metrdesmom}
\end{equation}
Using this metric we can define the invariant line-element in momentum-space as the geodesic distance from the momentum-space origin
\begin{equation}
d\mathit{k}^2(0,p)=\int_\lambda \tilde{\eta}^{\alpha\beta}(p)\dot{\lambda}_\alpha\dot{\lambda}_\beta\,ds={\cal C}(p)\,,
\end{equation}
in which $s$ is the variable with which we parametrise our geodesic $\lambda(s)$ connecting the point at $p$ in which a particle lies to the origin, and in which 
\begin{equation}
{\cal C}(p)=p_0^2-p^2-\ell p_0 p^2\,,\label{Casimir}
\end{equation}
where, by definition, in 2+1 dimensions $p^2=p_L^2+p_T^2$. Being the ${\cal C}(p)$ invariant, we can identify it as the Casimir operator of the deSitter momentum-space transformation generators algebra
\begin{eqnarray}
&\{p_0,p_i\}=0\;,\;\;\;\{p_i,p_j\}=0\;,\;\;\;
\{\mathcal{N}_{(i)},{\cal R}\}=
\epsilon_{ij}{\cal N}_{(j)}\;,\;\;\;\{{\cal N}_{(i)},{\cal N}_{(j)}\}=\epsilon_{ij}{\cal R}\,,&\\
&\{\mathcal{N}_{(i)},p_0\}=-p_i\;,\;\;\;\{\mathcal{N}_{(i)},p_j\}=-\delta_j^i\left(p_0-\ell p_0^2+\frac{\ell}{2}p^2\right)+\ell p_i p_j\,,&
\end{eqnarray}
in which the boost ${\cal N}_{(i)}$ and the rotation ${\cal R}$ generators can be represented in terms of the $\chi^\mu$ coordinates as
\begin{equation}
{\cal N}_{(i)}= \chi^0 p_i +\chi^i \left(p_0-\ell p_0^2+\frac{\ell}{2}p^2\right)\;,\;\;\;{\cal R}=\chi^L p_T - \chi^T p_L\,.\label{boost1}
\end{equation}
An important deSitter momentum-space feature to take into account is the  deformation of the symplectic structure between momenta and coordinates, given by the non trivial relation between the coordinate components (\ref{noncomm}). Therefore
\begin{equation}
\begin{split}
\{p_0,\chi^0\}=1\;\;&,\;\{p_0,\chi^j\}=0\\
\{p_i,\chi^0\}=-\ell p_i\;&,\;\;\{p_i,\chi^j\}=\delta_i^j\,.\label{symplchi}
\end{split}
\end{equation}
It is easy to check that (\ref{symplchi}) and (\ref{noncomm}) satisfy all Jacobi identities.\\
We can obtain the finite action of the boost transformation by means of the Poisson brackets of its generator ${\cal N}_{(i)}$ through the map
\begin{equation}
{\cal B}_{(i)}\rhd f(x,p)=f(x,p)-\xi\{{\cal N}_{(i)},f(x,p)\}+\frac{\xi^2}{2!}\{{\cal N}_{(i)},\{{\cal N}_{(i)},f(x,p)\}\}+...
\end{equation}
However in the following sections instead of summing all the $\xi^n$ contributes we will, for sake of simplicity, just integrate the first-order term of the series expansion.

\section{The boost parameters in Relative Locality}

In this first paragraph we will review some basic concepts of Relative Locality in 1+1D and, at the end, we will show how the special-relativistic parameter $\beta$ and $\gamma$ find a rather simple interpretation even in a curve-momentum-space framework. In Special Relativity, in order to identify the physical meaning of $\beta$ we take advantage from the mathematical relation between hyperbolic sine and cosine
\begin{equation}
 \cosh^2(\xi)-\sinh^2(\xi)=1\label{rel1}\,,
\end{equation}
then we re-define the two functions as
\begin{equation}
\cosh(\xi)=\gamma\;\;\;\;\;\;\sinh(\xi)=\beta\gamma\,,\label{hypbg}
\end{equation}
therefore (\ref{rel1}) determines the connection between the two parameters (which still have no physical interpretation for now)
\begin{equation}
 \gamma=\frac{1}{\sqrt{1-\beta^2}}.
\end{equation}
Of course those relations find a useful application in describing the coordinate and momenta transformations between two observers boosted one respect the other. We let the special-relativistic boost generator ${\cal N}_{SR}=x^1 p_0+x^0 p_1$ act on momenta $p_\alpha$ through the Poisson-bracket formalism. Then we find the infinitesimal variation of momentum-space coordinates with respect the {\it rapidity} parameter $\xi$: 
\begin{equation}
\left\{\begin{array}{ll}  
\frac{d p_0}{d\xi}=-\{{\cal N}_{SR},p_0\}=p_1 \\
\frac{dp_1}{d\xi}=-\{{\cal N}_{SR},p_1\}=p_0\label{relclass1}
\end{array}\right.
\end{equation}
System (\ref{relclass1}) can be easily solved for example using the {\it ab initio} conditions $p_0(0)=\mu$, $p_1(0)=0$. With this choice we find the usual \begin{equation}
p_0(\xi)=\mu\cosh(\xi)\;\;\;\;\;\;p_1(\xi)=\mu\sinh(\xi)\,.\label{p0ep1}
\end{equation}
Now from equation (\ref{rel1}) is straightforward to obtain the special-relativistic invariant dispersion relation
$$
p_0^2-p_1^2=\mu^2\,,
$$
and also the physical interpretation for the $\beta$ parameter:
$$
\beta=\frac{\sinh(\xi)}{\cosh(\xi)}=\frac{p_1}{p_0}\equiv v_1\,,
$$
which is the particle's velocity we find in the expression of special-relativistic wordlines.\\
In Relative Locality we proceed in a quite similar way. The RL version of (\ref{relclass1}) was already found in DSR literature \cite{trasfallord1,trasfallord2} and defined in a curved momentum-space framework in \cite{FlaGiu} at all orders in the deformation parameter $\ell$. It is sufficient for our purposes to discuss the first order expansion formalism which was already used to explore synchrotron radiation in Deformed Special Relativity \cite{synchrotron}. Indeed the transformations of our curve-momentum-space coordinates can be obtained from the deformed boost generator (see (\ref{boost1})) action: 
 \begin{equation}
  \left\{\begin{array}{ll} 
\frac{dp_0(\xi)}{d\xi}=-\{{\cal N},p_0\}=p_1(\xi)  \\
\frac{dp_1(\xi)}{d\xi}=-\{{\cal N},p_1\}=p_0(\xi)-\ell p_0^2(\xi)-\frac{\ell}{2}p_1^2(\xi)\label{rel3}
\end{array}\right.\,.
\end{equation}
This differential equation system can be solved by perturbing the solutions we found in the classical case (\ref{p0ep1}) as 
\begin{equation}
p_0(\xi)=\mu\cosh(\xi)+\ell a(\xi)\;,\;\;\; p_1(\xi)=\mu\sinh(\xi)+\ell b(\xi)\,.  \label{perturb1}
\end{equation}
Thus, using (\ref{perturb1}), we reduce (\ref{rel3}) to the relations
\begin{eqnarray}
 a(\xi)-\frac{d^2a(\xi)}{d\xi^2}&=&\mu^2\left(\cosh^2(\xi)+\frac{1}{2}\sinh^2(\xi)\right)\\
b(\xi)&=&\frac{da(\xi)}{d\xi}\,,
\end{eqnarray}
from which we finally obtain the solutions
\begin{eqnarray}
p_0(\xi)&=&\mu\cosh(\xi)-\ell \frac{\mu^2}{2}\sinh^2(\xi)\label{p01+1}\\    
p_1(\xi)&=&\mu\sinh(\xi)-\ell \mu^2\sinh(\xi)\cosh(\xi)\label{p11+1}\,.
\end{eqnarray}
We can now verify that if we assume the energy-momentum dispersion relation to be deformed according to (\ref{Casimir}) we still can obtain a coherent picture for the invariance of the particle mass, in fact since
\begin{equation}
\sinh(\xi) \simeq \frac{p_1}{\mu} (1+\ell\frac{p_0}{\mu}) \;,\;\;\;   \cosh(\xi) \simeq \frac{p_0}{\mu}+\frac{\ell}{2}\frac{p_1^2}{\mu}\,,
\end{equation}
we can again rely on (\ref{rel1}) (which is purely a relation between hyperbolic functions and then not model-dependent at all) to define our Modified Dispersion Relation (MDR), invariant under deformed boost transformations
$$
p_0^2-p_1^2-\ell p^2 p_0=\mu^2.
$$
We can now recover the generic definitions (\ref{hypbg}) for $\beta, \gamma$ (which as stated above still is not model-dependent). Therefore:
\begin{equation}
\beta=\tanh(\xi)=\frac{ p_1 (1+\ell p_0)}{ p_0+\frac{\ell}{2} p_1^2}=\frac{|p_1|}{\sqrt{p^2+\mu^2}}+\ell|p_1|\left(1-\frac{p_1^2}{p_1^2+\mu^2}\right)\label{beta}\,.
\end{equation}
This result is very important, since also in the $\ell$-deformed framework $\beta$ can be interpreted as the velocity of a boosted particle in the laboratory reference frame. We can in fact notice that relation (\ref{beta}) is exactly the coordinate velocity found in previous Relative Locality works \cite{bob,kbob}. Therefore we can still express $\beta=v_1/c$, although its dependence on momenta is unavoidably non trivial.\\
It is important for phenomenological purposes to notice that the same thing is not true anymore in a symmetry-breakdown scenario, since, in this case we do not modify (\ref{relclass1}), and then the relation between $\beta_{LIV}$ and wordline velocity is not trivial anymore: $\beta_{LIV}= v_1(1-\ell p_0)$. The possibility of having a departure from the identification between the $\beta$ parameter and velocity $v$ is usually not taken into account in Lorentz-invariance violation literature (see {\it exempli gratia} \cite{Kostelecky1,GlashowOpera}), and maybe should be better deepened.

\subsection{Deformed Lorentz momenta transformations in 2+1 dimensions}

It is maybe important to deepen our exploration on Relative Locality with deSitter momentum-space in more than one spatial dimension, since it shows a peculiar feature which in literature is called {\it Transverse Relative Locality} \cite{bob,leelaurentGRB,transverse,transverseproc}. This feature is an important aspect of theories with relativity of locality since it provides interesting phenomenological effects as we will see further.
In 2+1 D the system of differential equations (\ref{rel3}) is enriched by a transverse-component equation:
\begin{equation}
\left\{\begin{array}{lll}
\frac{dp_0(\xi)}{d\xi}=-\{{\cal N}_{(L)},p_0\}=p_L(\xi)  \\
\frac{dp_L(\xi)}{d\xi}=-\{{\cal N}_{(L)},p_L\}=p_0(\xi)-\ell p_0^2(\xi)+\frac{\ell}{2}|p|^2(\xi)-\ell p_L^2\\
\frac{dp_T(\xi)}{d\xi}=-\{{\cal N}_{(L)},p_T\}=-\ell p_L p_T
\end{array}\right.\label{sistemap}
\end{equation} 
We can solve the system perturbatively as done in the previous section with system (\ref{rel3}), fixing the generic {\it ab initio} conditions $p_0(0)=\bar{p}_0$, $p_L(0)=\bar{p}_L$ and $p_T(0)=\bar{p}_T$; given those we find the generic solutions 
\begin{eqnarray}
p_0(\xi)&=&\bar{p}_0\cosh(\xi)+\bar{p}_L\sinh(\xi)-\frac{\ell}{2}(\cosh(\xi)-1) \left( \bar{p}_0^2(\cosh(\xi)+1) + \bar{p}_L^2\cosh(\xi)- \bar{p}_T^2 + 2\bar{p}_0\bar{p}_L\sinh(\xi) \right)\,,\label{p02+1}\\
p_L(\xi)&=&\bar{p}_L\cosh(\xi)+\bar{p}_0\sinh(\xi)+\ell\left( \bar{p}_0\bar{p}_L(1+\cosh(\xi)-2\cosh^2(\xi))+\frac{1}{2}|\bar{p}|^2\sinh(\xi) -(\bar{p}_0^2+\bar{p}_L^2)\sinh(\xi)\cosh(\xi)\right)\,,\label{pL2+1}\\
p_T(\xi)&=&\bar{p}_T+\ell\bar{p}_T\left( \bar{p}_0(1-\cosh(\xi))-\bar{p}_L\sinh(\xi)\right)\,.\label{pT2+1}
\end{eqnarray}
It is very easy to verify that those solutions reduces to (\ref{p01+1}) and (\ref{p11+1}) if we fix the initial conditions as $\bar{p}_0=\mu$, $\bar{p}_L=0$ and $\bar{p}_T=0$. Another important property of solutions (\ref{p02+1}), (\ref{pL2+1}) and (\ref{pT2+1}) is that they verify the invariance of the deformed dispersion relation defined by the Casimir (\ref{Casimir}) at all orders in $\xi$, in fact we observe that
\begin{equation}
p_0^2(\xi)-(p_L^2(\xi)+p_T^2(\xi))-\ell p_0(\xi)(p_L^2(\xi)+p_T^2(\xi))=\bar{p}_0^2-(\bar{p}_L^2+\bar{p}_T^2)-\ell \bar{p}_0(\bar{p}_L^2+\bar{p}_T^2)=\mu^2\,,\label{massinv}
\end{equation}
as we could expect, given relation $\{{\cal N}_{(i)},\mathcal{C}\}=0$. One thing we can notice from equation (\ref{massinv}) is that the invariance of the dispersion relation is strictly related to the transformations of all the components of momenta. While in SR the transformation of the $p_0$ and the $p_L$ components compensate each other (where $L$ is chosen as the boost direction), in RL we need to take into account also the transverse one to ensure the invariance of the MDR. Since (\ref{pL2+1}) and (\ref{pT2+1}) balance each other harmoniously, there is no point in studying the evolution of the angle $\theta=\arctan(p_T(\beta)/p_L(\beta))$ between the two momenta (we would obtain a practically indistinguishable behaviour from the SR one). On the other hand it may be of some interest to analyse the behaviour of the single momentum components.  
\begin{figure}[h]
\centering
\includegraphics[scale=0.83]{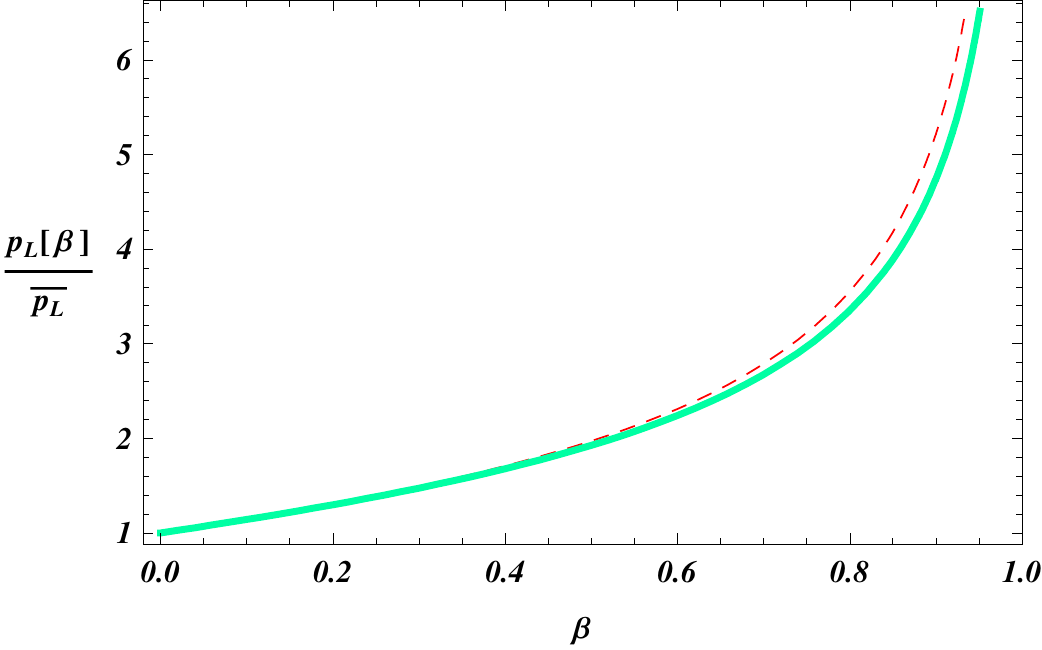}
\includegraphics[scale=0.85]{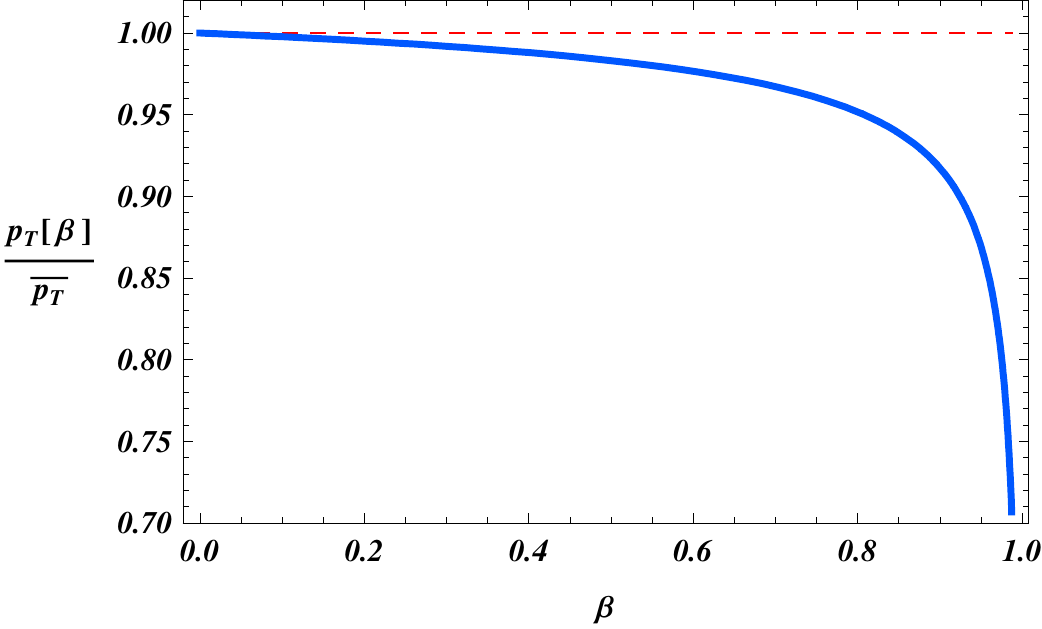}
\caption{\footnotesize In those pictures we represent the behaviour respectively of the $p_L$ and the $p_T$ component of momenta, for different values of the $\beta$ parameter. The straight lines obey to (\ref{pL2+1}) and (\ref{pT2+1}) transformation laws, while the dashed ones represent the special relativistic case. Of course in order to show explicitly the differences between those two theories, the momenta absolute value has been fixed at some consistent fraction of our deformation scale ($|p|\sim 0.03\,\ell^{-1}$).}
\label{fig:LvsT}
\end{figure}

While $p_L(\beta)$ basically follows the special relativistic curve, $p_T(\beta)$ shows a sensitively different behaviour than the SR case, at some orders of magnitude below our deformation scale $\ell$. This could be an important feature for further phenomenological investigations of Relative Locality, for example for what concerns the study of deformed particle vertices. We will not deepen those aspects in this paper for they might deserve dedicated studies, instead we are here more interested in characterising deformed Lorentz-transformations effects also for space-time.

\section{Coordinate transformations and Rainbow metrics} 

In literature many studies try to define the behaviour of Relative Locality in presence of spacetime curvature \cite{KowaJack1,KowaJack2}. In order to support those efforts it may be of some interest to develop a phenomenology of RL effects for example at a cosmological scale. An important mathematical tool which could be very useful in this kind of analysis is the Rainbow metrics formalism \cite{SmolinRainbow}. In this paper we have now the possibility to suggest how those momentum-dependent metrics should naturally arise in the Minkowskian limit of Relative Locality. In fact, as done for momenta (\ref{sistemap}), we can define the different spacetime coordinatizations, that two boosted observers would use to describe physical phenomena, by solving the system 
\begin{equation}
\left\{\begin{array}{lll}
\frac{d\chi^0(\xi)}{d\xi}=-\{{\cal N}_{(L)},\chi^0\}=-\chi^L(\xi)+\ell\left(\chi^L(\xi) p_0(\xi)+\chi^0(\xi) p_L(\xi)\right)  \\
\frac{d\chi^L(\xi)}{d\xi}=-\{{\cal N}_{(L)},\chi^L\}=-\chi^0(\xi)\\
\frac{d\chi^T(\xi)}{d\xi}=-\{{\cal N}_{(L)},\chi^T\}=-\ell \left( \chi^L(\xi) p_T(\xi)-\chi^T(\xi) p_L(\xi) \right)
\end{array}\right.\label{siscoord}
\end{equation}
As usual we opt for solving (\ref{siscoord}) perturbatively at first order in $\ell$, using the solutions we found in last section (\ref{p02+1}), (\ref{pL2+1}) and (\ref{pT2+1}), to write the explicit expressions for momenta $p_\mu(\xi)$. The solutions of system (\ref{siscoord}) for generic {\it ab initio} conditions $\chi^\mu(0)=\bar{\chi}^\mu$ are 
\begin{eqnarray}
\chi^0(\xi)&=&\bar{\chi}^0\cosh(\xi)-\bar{\chi}^L\sinh(\xi)+\ell\sinh(\xi)\left(\bar{\chi}^L\bar{p}_0+\bar{\chi}^0\bar{p}_L\right)\,,\label{chi02+1}\\
\chi^L(\xi)&=&\bar{\chi}^L\cosh(\xi)-\bar{\chi}^0\sinh(\xi)+\ell(1-\cosh(\xi))\left(\bar{\chi}^L\bar{p}_0+\bar{\chi}^0\bar{p}_L\right)\,,\label{chiL2+1}\\
\chi^T(\xi)&=&\bar{\chi}^T + \ell \left( (\cosh(\xi)-1) \left(\bar{\chi}^T \bar{p}_0 + \bar{\chi}^0 \bar{p}_T\right) - \sinh(\xi)\left(\bar{\chi}^L\bar{p}_T-\bar{\chi}^T\bar{p}_L\right) \right)\,.\label{chiT2+1}
\end{eqnarray}
Those solution can help us to define the Relative-Locality-invariant line element $ds^2$ at all orders in $\xi$, in the same exact way we showed the invariance of the dispersion relation in (\ref{massinv}). We can therefore observe that two boosted observers will agree on
\begin{equation}
\chi^0(\xi)^2 - \left(\chi^L(\xi)^2+\chi^T(\xi)^2\right)(1-2\ell p_0(\xi)) + 2\ell \chi^0(\xi)\chi^i(\xi)p_i(\xi) = 
\left(\bar{\chi}^0\right)^2 - \left(\left(\bar{\chi}^L \right)^2+\left(\bar{\chi}^T \right)^2\right)(1-2\ell \bar{p}_0) + 2\ell \bar{\chi}^0 \bar{\chi}^i \bar{p}_i \,.\label{linelinv}
\end{equation}  
Relation (\ref{linelinv}) can be more synthetically expressed through a metric formalism as
\begin{equation}
\Delta s^2=\tilde{\eta}_{\mu\nu}^{(\chi)}(p)\chi^\mu \chi^\nu\,,
\end{equation}
where the momentum-dependent 2+1D Minkowskian metric $\tilde{\eta}$ is defined as
\begin{equation}
\tilde{\eta}_{\mu\nu}^{(\chi)}(p)=\left(
\begin{array}{ccc}
1 & \ell p_L & \ell p_T\\
\ell p_L & -(1-2\ell p_0) & 0\\
\ell p_T & 0 & -(1-2\ell p_0)
\end{array}
\right)\,.\label{Rainchi}
\end{equation}
This example shows explicitly how the Rainbow metrics formalism is naturally implemented in the Relative Locality theory. The main difference between the Rainbow formalism used in \cite{SmolinRainbow} and the one we show in this paper is that in Relative Locality the definition of metric $\tilde{\eta}$ is not obtained through the modified dispersion relation as $m^2=g^{\alpha\beta}_{(R)}(p)p_\alpha p_\beta$. {\it Vice versa} in RL both MDR and spacetime Rainbow metric are shaped on the curve momentum-space metric (\ref{metrdesmom}).\\
It may seem that metric (\ref{Rainchi}) may not be dual to momentum-space metric (\ref{metrdesmom}), because of the off-diagonal elements. That's not a problem, because we don't expect metric $\tilde{\eta}^{(\chi)}_{\alpha\beta}$ to be dual to the momentum-space one, since noncommutative coordinates $\chi^\mu$ have a non-trivial symplectic sector (\ref{symplchi}). Duality is instead required for commutative coordinates $x^\beta$ which satisfy $\{p_\alpha,x^\beta\}=\delta_\alpha^\beta$. The liaison between $\chi^\alpha$ and $x^\beta$ coordinates is very-well known in Relative Locality literature \cite{principle,FlaGiu,kbob} and is
\begin{equation}
\chi^\alpha=\tau^\alpha_\beta (p) x^\beta=(\delta^\alpha_\beta-\ell\delta^\alpha_0\delta_\beta^j p_j)x^\beta\,,\label{noncommvscomm}
\end{equation} 
where the $\tau^\alpha_\beta (p)$ are the translation deSitter momentum-space killing vectors (see \cite{lateshift} for a clear discussion of the physical implications of this feature). \\
Using relation (\ref{noncommvscomm}) we can find that
\begin{equation}
\Delta s^2=\tilde{\eta}_{\mu\nu}^{(\chi)}(p)\chi^\mu \chi^\nu= \tilde{\eta}_{\mu\nu}(p) x^\mu x^\nu\,,
\end{equation}
where
\begin{equation}
\tilde{\eta}_{\mu\nu}(p)=\left(
\begin{array}{ccc}
1 & 0 & 0\\
0 & -(1-2\ell p_0) & 0\\
0 & 0 & -(1-2\ell p_0)
\end{array}
\right)\,.\label{Rainx}
\end{equation}
Then, confronting (\ref{Rainx}) with (\ref{metrdesmom}) it is now clear how duality between spacetime and momentum-space metrics is manifest, since $\tilde{\eta}^{\alpha\gamma}\tilde{\eta}_{\gamma\beta}=\delta^\alpha_\beta$.

\subsection{Clocks and transverse effects}

\begin{figure}[h!]
\centering
\fbox{\includegraphics[scale=0.2]{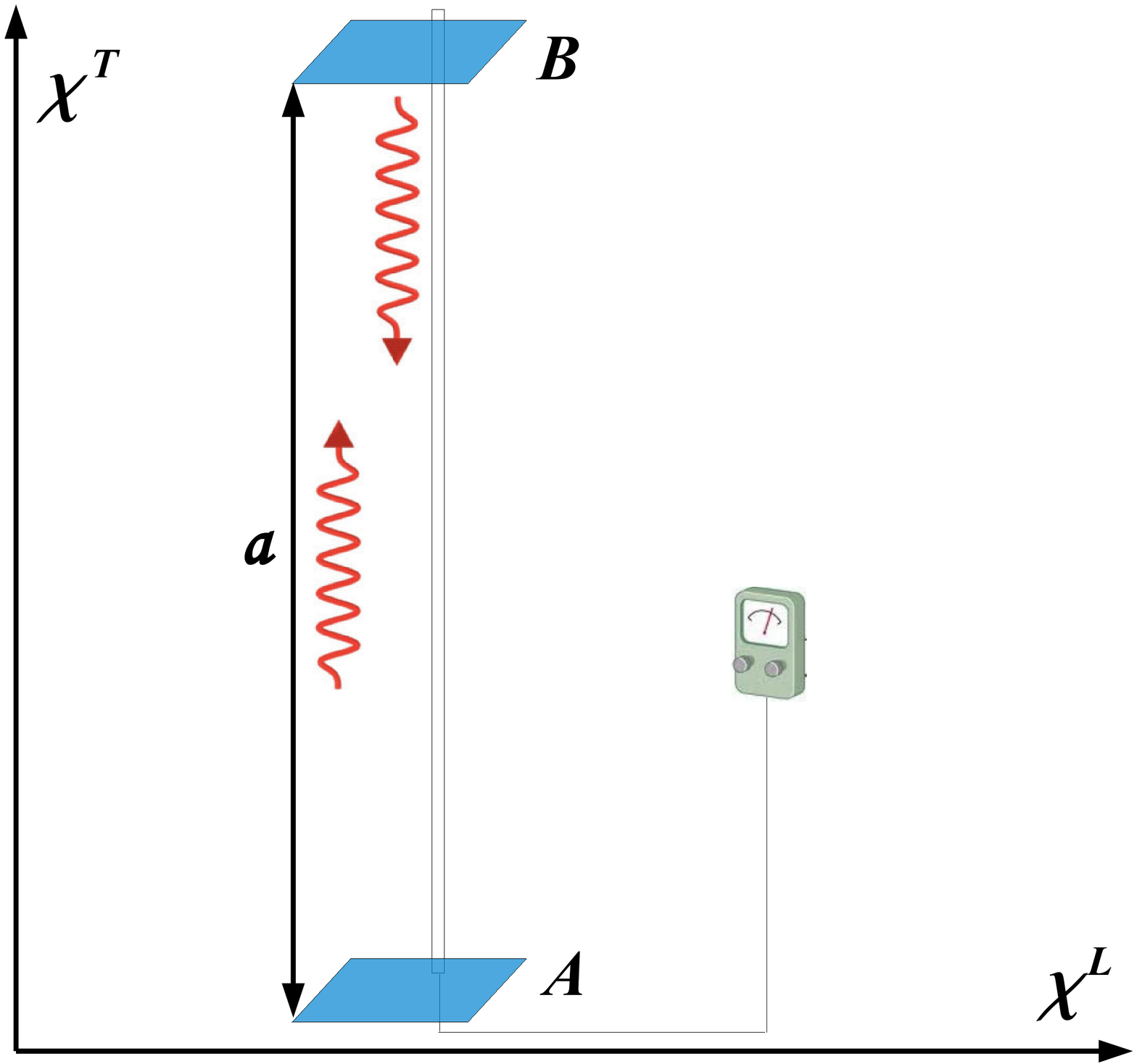}}
\caption{\footnotesize Einstein synchronization convention: a photon is sent from the emission point {\it A} at time $\chi^0=0$ thowards the mirror in {\it B}. The following detection of the reflected photon in {\it A} gives us the definition of time.}
\label{fig:Clock}
\end{figure}

In order to explore special Relative Locality phenomenology we should now define the procedure we use to identify what we call time intervals. As in usual Special Relativity in RL we can rely on the absoluteness of the speed of light using an Einstein clock of length $a$ (see figure \ref{fig:Clock}) to define time units. The only problem we should be careful about is the non trivial relation between lengths and time intervals. 

In fact, while according to (\ref{beta}) in $\chi^\alpha$ coordinates our photons have trivial wordlines
\begin{equation}
\chi^T-\bar{\chi}^T=-\left(\chi^0-\bar{\chi}^0\right)\,,\label{wordlichi}
\end{equation}
on the other hand we have deformed translations \cite{kbob,lateshift} due to the non trivial symplectic sector (\ref{symplchi}). Then the ideal interaction point between a photon emitted in {\it A} and the mirror in {\it B} has coordinates
\begin{equation}
\chi_{(B)}^\nu=\bar{\chi}_{(A)}^\nu-a^\mu\{p_\mu,\chi_{(A)}^\nu\}=\bar{\chi}_{(A)}^\nu- a^\mu(\delta_\mu^\nu-\ell\delta_0^\nu\delta_\mu^i \bar{p}_i)\,,\label{traslchiB}
\end{equation}
where with ($\bar{\chi}^L_{(A)}=0,\bar{\chi}^T_{(A)}=0$) we indicate the emission point coordinates in the {\it A} frame. Then, using (\ref{traslchiB}) with (\ref{wordlichi}) we obtain that according to the translated observer (whose spatial origin is in $(\chi^L_{(B)}=0,\chi^T_{(B)}=0)$) the emission point has coordinates
\begin{equation}
\bar{\chi}^0_{(B)}=-a^0 + \ell a^T \bar{p}_T\;,\;\;\;\bar{\chi}^L_{(B)}=0\;,\;\;\;\bar{\chi}^T_{(B)}=-a^T\,,
\end{equation}
Then the observer in {\it B} infers different emission times for different photon energies. Moreover, using the wordline expression (\ref{wordlichi}) we can verify that also the photon time-of-arrival at mirror in {\it B} is momentum dependent:
\begin{equation}
\chi^0_{(B)}(\chi^L_{(B)}=0,\chi^T_{(B)}=a)=a^0-a^T-\ell a^T \bar{p}_T\,.\label{riflesso}
\end{equation}
All this may result a little bit weird to a reader facing Relative Locality-related effects for the first time, but we should keep in mind that all those features are merely a coordinate artifact, due to the curvature of momentum-space. This concept is even clearer when one clarifies what to expect from the entire emission-reflection-detection process.
In fact since the detector is placed in {\it A}, we can check if such a momentum-dependency is still present in the time interval measured by our device by calculating where the observer in {\it A} would infer the emission point. First of all we have to fix, using the inverse of transformations (\ref{traslchiB}), how {\it A} would express the photon reflection point (\ref{riflesso}):
\begin{equation}
\chi_{(A)}^\nu=\bar{\chi}_{(B)}^\nu+a^\mu\{p_\mu,\chi_{(B)}^\nu\}=\bar{\chi}_{(B)}^\nu+ a^\mu(\delta_\mu^\nu-\ell\delta_0^\nu\delta_\mu^i \bar{p}_i)\,.\label{traslchiA}
\end{equation}
Then, setting the result of (\ref{traslchiA}) as starting point for the wordlines (\ref{wordlichi}), and considering that momentum $p_T$ now points in the opposite direction, we obtain that the observer in {\it A} infers the emission time to be
\begin{equation}
\bar{\chi}_{(A)}^0=-2 a^T-2\ell a^T \bar{p}_T\,.
\end{equation}
Therefore, as we expected, since the momentum-dependence of the photon time of flight that {\it B} observes is a physical effect (\ref{traslchiB}), we obtain that the time interval definition in Relative Locality depends explicitly on momentum-space curvature:
\begin{equation}
\Delta\chi^0\simeq 2 a(1-\ell \bar{p}_0)\,.
\end{equation}
The reason why we have formalised our theory using coordinates with apparently complicated relations between each other (\ref{noncomm}) and non trivial symplectic sector (\ref{symplchi}), is that we have been able to express the physical effect just as a feature of the deformed translations. If instead of using the $\chi^\alpha$ coordinates we had used the commutative $x^\alpha$ ones, we would have payed the simplification of the mathematical formalism with a more complex description of the whole synchronization mechanism (though the physical result would have been the same).\\ 
Using this coordinatization, it is now easy to obtain the time-interval expression for a boosted observer. In fact, if we imagine to observe the device in Figure \ref{fig:Clock} from a reference frame boosted along the $\chi^L$ direction, since any transverse effect on momentum is suppressed by a factor ${\cal O}(\ell^2)$, according to (\ref{chi02+1}) we would define the time interval just as
\begin{equation}
\Delta\chi^0(\beta)=2a\gamma (1-\ell \bar{p}_0)\,,\label{pretimeinterv}
\end{equation}
where $a$ is defined in the rest frame. However a boosted observer would not express the clock length in terms of $a$. If instead we wish to express our time-interval in terms of the boosted reference frame observables, we should take into account also the relatively local transverse effect. Then, being ${\cal L}_{(a)}$ the clock length measured by the boosted observer, using (\ref{chiT2+1}), equation (\ref{pretimeinterv}) becomes
\begin{equation}
\Delta\chi^0(\beta)=2\mathcal{L}_{(a)}\gamma(1-\ell (2\gamma -1 +\beta\gamma )\bar{p}_0 )\,.
\end{equation}
In order to imagine a way to detect this effect, we can borrow a common idea in quantum-gravity literature, considering the time-delay of two simultaneously-emitted photons carrying different energies \cite{phenomenology} in two different boosted reference-frames. While in the clock's reference-frame we expect a momentum-dependent time-delay only amplified from the size of length $a$, on the other hand, according to a boosted observer, the two photons should reach the detector at different times, whose difference for $\gamma\gg 1$ is $\delta T\sim\ell\gamma^2{\cal L}_a\delta E$. In order for this effect to have any significance, an ideal {\it gedanken experiment} based on it should then compare the observations of two boosted observers with high boost parameter $\gamma$, for the ricochet of two photons with big energy difference $\delta E$, in a clock with large $\mathcal{L}_a$, to compensate the tiny value of $\ell$.

\section{Closing remarks}

In Quantum-Gravity phenomenology it is always complex to define observables and consequently to fix upper bounds to the parameters we use to formalise the effects. It is then, in my experience, useful to express those effects as corrections to the classical models. This is precisely the spirit of this whole article in which the manifestations of momentum-space curvature are expressed as a deformation of Lorentz transformations, modelized in terms of the usual $\beta$ and $\gamma$ parameters. With this formalisation it is pretty simple to characterise the deformation effects, even the most unexpected ones, like the boost-related Transverse Relative Locality. About this rather unexplored scenario of transverse effects in deSitter momentum-space, it may be interesting to verify if such features can be of some help in identifying an upper limit for phenomenological parameters. For example for analysis such like the one reported in \cite{neutrini}, for which is crucial the identification of the origin point of detected particles.\\
Is also interesting for phenomenological purposes the discussion about the deformed (momentum-dependent) law for time-intervals dilatation (the boost parameter $\gamma$ appears to act like a magnifier for RL-effects), and it might require a dedicated research programme to identify the most promising applications that might allow to unveil such effects. But the payout that could be expected appears to be worth the effort, since such a novel window on the Planck-scale realm could have particularly significant impact on our ability to investigate the quantum-gravity problem.\\ 
Also for what concerns the more academic/conceptual side of the issues here discussed, these studies should motivate further investigation, particularly for what concerns the identification of a characteristic metric formalism for Relative Locality which could also be extremely important from the phenomenological side. As discussed in Section III.\\
\\
{\it This work is supported by a "La Sapienza" fellowship (perfezionamento all'estero, area CUN1). It was also made possible
in part through the generous hospitality of the Perimeter Institute for Theoretical Physics. I acknowledge useful conversation with Dr. Giacomo Rosati and Dr. Flavio Mercati, who I also thank for his great noodles.}

\end{document}